\documentclass[preprints,article,accept,pdftex,moreauthors]{Definitions/mdpi} 
\firstpage{1} 
\pubvolume{1}
\issuenum{1}
\articlenumber{0}
\pubyear{2025}
\copyrightyear{2025}
\datereceived{ } 
\daterevised{ } 
\dateaccepted{ } 
\datepublished{ } 
\hreflink{https://doi.org/} 

\Title{Profiling a Rydberg-Atom Electric Field Sensor for Off-Resonant Detection of Sub-100~MHz RF Signals}

\TitleCitation{Title}

\Author{Michael A. Viray*\orcidA, Abby Halasi-Kun\orcidC, Baran N. Kayim \orcidD, Brian C. Sawyer\orcidE, Robert Wyllie\orcidF, and David S. La~Mantia$^{ \dagger}$\orcidB}

\AuthorNames{Firstname Lastname, Firstname Lastname and Firstname Lastname}

\isAPAStyle{%
       \AuthorCitation{Viray, F., Lastname, F., \& Lastname, F.}
         }{%
        \isChicagoStyle{%
        \AuthorCitation{Viray, Firstname, Firstname Lastname, and Firstname Lastname.}
        }{
        \AuthorCitation{Viray, M. A., Halasi-Kun, A., Kayim, B.N.,  Sawyer, B.C., Wyllie, R., La~Mantia, D.S.}
        }
}

\address{%
\quad Georgia Tech Research Institute, Atlanta, GA 30332, USA
}

\corres{Correspondence: michael.viray@gtri.gatech.edu}

\firstnote{Correspondence: david.lamantia@gtri.gatech.edu}

\abstract{We present a Rydberg-Atom electric field sensor optimized to detect signals at sub-100~MHz carrier frequencies. The sensing setup employs a sapphire vapor cell that allows for detection of signals below 100~MHz \textemdash typical vapor cells made of glass or quartz demonstrate strong screening of radio frequency (RF) signals in this frequency regime. Applied signals are detected by observing AC~Stark shifts in the atomic vapor energy levels. As a test case for the commercial utility of this receiver, we perform our tests at several carrier frequencies in the Industrial, Scientific, and Medical (ISM) band. At each carrier frequency, we report sensitivity, minimum detectable field, and detectable electric-field dynamic range. We also present a routine for optimizing off-resonant signal detection by tuning experimental parameters such as Rydberg coupler laser detuning and RF local oscillator strength. This Python-based optimization routine, which can be used at any off-resonant carrier frequency, is shared on Github for others to use in their own investigations.}


\begin{document}




\section{Introduction}
\label{sec:introduction}

Rydberg atom-based electric field sensing is a burgeoning field with great potential for practical applications. Atoms in Rydberg states have exaggerated physical properties such as transition dipole moments and polarizabilities \citep{gallagherbook}, which make them highly responsive to external electric fields. The underpinning phenomenon behind Rydberg field sensing is electromagnetically induced transparency~(EIT), in which certain configurations of excitation lasers and atomic states produce narrow spectroscopic features within absorption windows \citep{harris1990, boller1991observation, boydbook}. Rydberg-atom spectroscopic lines respond in well-understood ways to external fields, and it is this response that allows Rydberg atom systems to act as sensors for RF signals. Furthermore, Rydberg atom systems are self-calibrated and fully SI-traceable, unlike classical antennas \citep{holloway2014, holloway2017atom}. Most Rydberg field sensing experiments make use of two-photon EIT, in which two lasers are used to excite atoms to states with high-principal quantum number ($n\sim 50$) states either zero ($nS$) or two ($nD$) quanta of orbital angular momentum due to selection rules. Resonant transitions from these initial states to other nearby Rydberg states are generally in the microwave frequency regime, and many experiments using two-photon EIT to measure electric field properties such as amplitude \citep{sedlacek2012}, phase \citep{simons2019}, and polarization \citep{sedlacek2013, ma2022electrodes} are aimed at these carrier frequencies. Subsequent versions of two-photon Rydberg field sensing have incorporated a local oscillator~(LO) field which, when combined with the signal field, produces a detectable optical heterodyne beat note. This method of detection yields superior metrics for minimum detectable field and sensitivity \citep{jing2020heterodyne} compared to direct atomic detection of the drive frequency, and it also provides a phase reference for angle of arrival measurements \citep{robinson2021aoa}. Aside from microwave-band detection, two-photon EIT has been used to measure electric fields from photoillumination \citep{ma2020dc} and plasmas \citep{anderson2017plasma}.

While two-photon EIT systems remain the most widely used in electric field sensing experiments, three-photon EIT systems have recently seen regular use as well \citep{carr2012threephoton, thaicharoen2019threephoton, duspayev2024threephoton, prajapati2023comparison, venu2025, prajapati2024threephoton}. Three-photon EIT systems offer certain advantages over two-photon systems. From a physics perspective, three-photon systems offer access to $nF$ Rydberg states. These states then allow access to $nF \rightarrow nG$ transitions, which are resonantly driven by UHF-band tones \citep{brown2023vhf, rotunno2023}. For a given $n$, these states also have larger polarizabilities than $S$ or $P$ states, making them advantageous for detection of off-resonant AC and DC fields \citep{kayim2026}. Thus, three-photon systems are a likely choice for UHF band (and below) detection. From an engineering standpoint, the three excitation wavelengths in three-photon systems can often be accessed with inexpensive, compact diode lasers. Conversely, the wavelength coupling the low-lying states to the Rydberg states (generally styled the 'coupler') in most two-photon schemes is not achievable through a convenient, off-the-shelf direct high-power diode and must be produced via a frequency-doubling system. These systems are cumbersome and less conducive for field deployment. The low-frequency detection capabilities and low-SWaP overall make three-photon EIT a more viable platform than two-photon EIT for eventual commercial applications. However, one technical complication is that three-photon systems expand the tunable parameter space (laser detuning, power, and propagation direction), making performance optimization challenging.

In this paper, we present a three-photon electric field sensing setup specifically designed to detect low-frequency RF signals. A sapphire vapor cell houses the rubidium atoms used for sensing, which has been shown to be more transparent to incoming low-frequency signals compared to vapor cells made of quartz or borosilicate glass \citep{jau2020vapor}. We apply the signal field using a custom built waveguide which can deliver single-mode signals from 0 to 300~MHz \citep{danielTEMpaper}. In addition to experimental hardware, we have also developed Python-based automation routines to calibrate the vapor cell/waveguide ensemble, optimizing parameters such as coupler laser detuning and LO strength and measuring the sensitivity with heterodyne beat note detection. These automation scripts allow for quick parameter optimization and data collection. In Sec.~\ref{sec:materials}, we give a thorough explanation of the experimental setup. In Sec.~\ref{sec:results}, we go into detail about the analysis scripts, sharing results from measurements at four different carrier frequencies: 6.78~MHz, 13.56~MHz, 27.12~MHz, and 40.68~MHz. These four carriers are the center frequencies of RF bands designated for industrial, scientific, and medical (ISM) applications \citep{fccTable}.

\section{Materials and Methods}
\label{sec:materials}

We employ a three-photon rubidium-85 EIT scheme\cite{brown2023vhf}. A 780~nm laser (probe) drives the $5S_{1/2}, F=3  \rightarrow 5P_{3/2}, F'=4$ transition, a 776~nm laser (dressing) drives the $5P_{3/2}, F'=4 \rightarrow 5D_{5/2}, F''=5$ transition, and a 1260~nm laser (coupler) drives the $5D_{5/2}, F''=5 \rightarrow 55F_{7/2}$ transition. All transitions excepting the final Rydberg state are hyperfine-resolved.  The lasers are frequency-locked to a commercial wavelength wavemeter, which itself is referenced to a separate laser locked to the D2 line in Rb. The 780~nm and 776~nm lasers are frequency-locked directly on resonance to their corresponding transition frequencies. The 1260~nm laser is locked 1~GHz red-detuned from the $5D_{5/2} \rightarrow 55F_{7/2}$ transition frequency, and the laser light passes through a fiber-coupled electro-optic modulator~(EOM) on its way to the vapor cell. The EOM is driven at $~$1~GHz by a GPS-referenced frequency synthesizer, and the +1~sideband compensates for the red-detuning imparted by the frequency lock (the fundamental and -1~sideband of the EOM are far-detuned from the Rydberg transition, and the AC Stark shift induced by these tones is within the linewidth of the Rydberg feature). Furthermore, the EOM drive frequency can be frequency-modulated ($\pm$90~MHz here), allowing us to linearly sweep across the EIT resonance with the +1~sideband. The higher-order sidebands (+2 and beyond) are either negligibly small or far-detuned from any upper Rydberg resonances.  This scheme for the 1260~nm laser allows us to easily switch between a well-calibrated scanning mode and a stable locked mode, both of which are critical for parameter optimization.

\begin{figure}
    \centering
    \includegraphics[width=\linewidth]{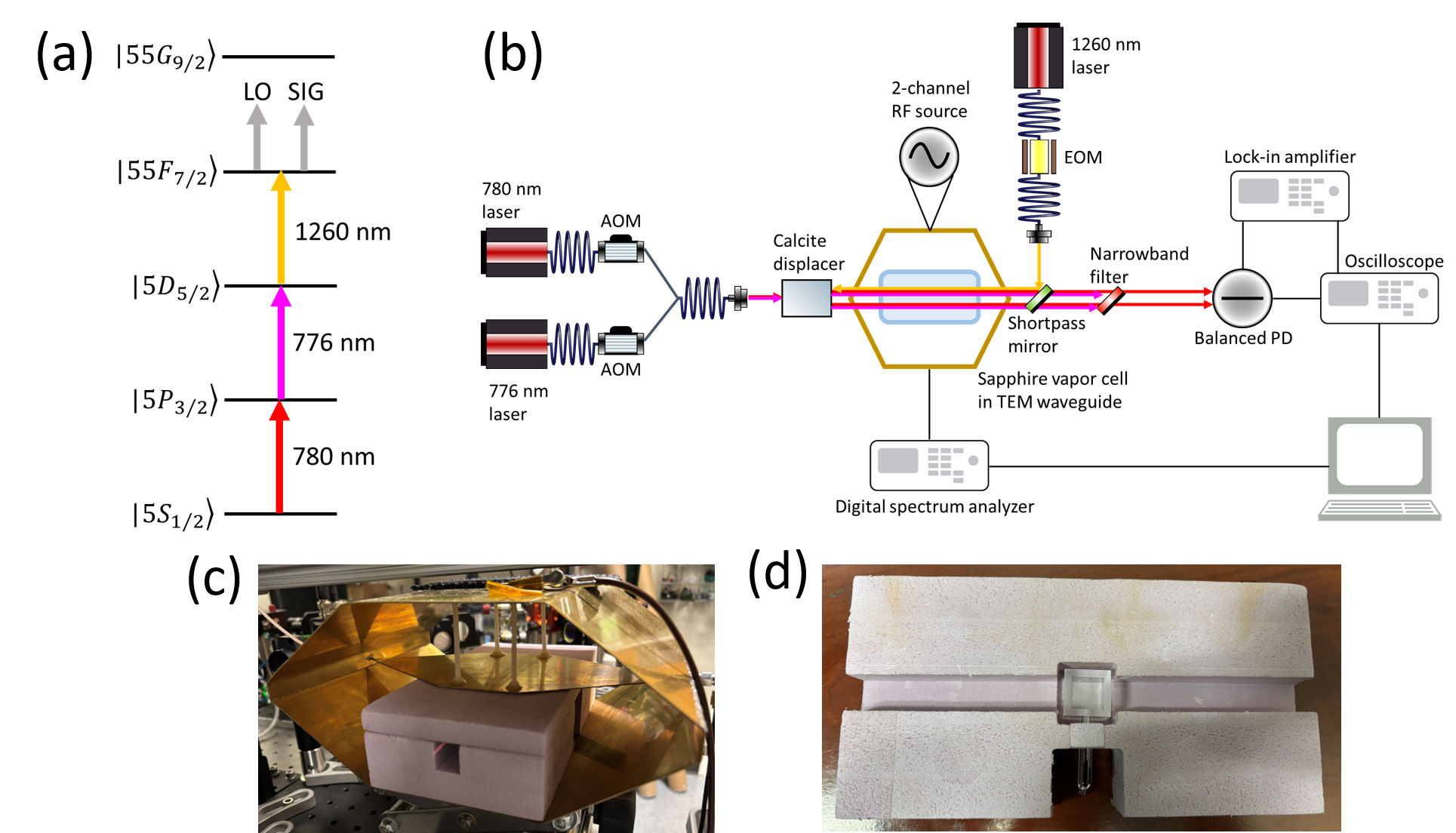}
    \caption{(a) Three-photon promotion scheme in Rb. (b) Illustration of experimental setup. (c) Brass TEM waveguide with the vapor cell foam enclosure inside. (d) The foam enclosure removed from the waveguide and opened, revealing the sapphire vapor cell.}
    \label{fig:materials}
\end{figure}

As mentioned in Sec.~\ref{sec:introduction}, we use a sapphire vapor cell for electric field sensing. The cell is fabricated by Japan Cell, and it is filled with natural abundance rubidium vapor. The cell is mounted in a Owens Corning Foamular enclosure (Fig.~\ref{fig:materials}d), which is transparent to these radio frequencies at our level of sensitivity. The enclosure is placed inside the brass coaxial waveguide, which delivers LO and signal fields to the vapor cell. The probe and dressing beams co-propagate through the vapor cell, while the coupler propagates in the opposite direction. We have found that counter-propagating the probe and dressing while both lasers are on-resonance results in unwanted blue light (420~nm) fluorescence and a noisier EIT signal (versus co-propagating); this problem is avoided when the beams are co-propagated. The probe and dressing lasers are also sent through a calcite displacer before passing through the cell, splitting the light into two paths with orthogonal linear polarizations, separated by 4~mm. The coupler laser overlaps with only one of the probe/dressing beams, thereby inducing EIT in one of the two probe laser arms. After passing through the cell, the probe and dressing beam pairs are dichroically separated, and the probe beams are sent into a differential photodiode for measurement. The experimental layout is illustrated in Fig.~\ref{fig:materials}b.

For each carrier frequency we observe with our setup, there are three distinct measurement phases, each with their own automated data collection and analysis scripts:

\begin{enumerate}
    \item Electric field calibration
    \item LO and coupler parameter optimization
    \item Sensitivity measurement
\end{enumerate}

In the first phase, the applied electric field from the waveguide is calibrated with direct atomic measurements. For constant carrier frequency, we apply a range of electric fields at different RF input powers, observe the resulting AC~Stark shift of the EIT resonance signal, and calculate the rms field strength by fitting the data to the following equation:
\begin{equation}
    \Delta f = \frac{1}{2} \alpha_\text{dyn} E_\text{rms}^2
    \label{eq:ACStarkShift}
\end{equation}

Here, $\Delta f$ is the measured frequency shift of the EIT line, $\alpha_\text{dyn}$ is the frequency-specific dynamic polarizability, and $E_\text{rms}$ is the rms electric field strength sensed by the atoms. The dynamic polarizability for each carrier frequency at fixed Rydberg state of $55F_{7/2}$ is calculated using the Alkali Rydberg Calculator \citep{vsibalic2017arc}. To aid in tracking the AC Stark shift, the coupler laser frequency is dithered with a $\sim$50 kHz current modulation as it scans, and the probe signal is demodulated with a lock-in amplifier referenced to the coupler modulation frequency. The output signal maps the EIT line to a dispersion curve and the peak position to the dispersion curve's zero crossing, which is simpler to monitor for tracking frequency shifts (see Fig.~\ref{fig:dither}).

As an added feature of the waveguide, the electric field strength can be reliably estimated in a fully classical manner by measuring the RF throughput on the exit port of the waveguide. These power measurements can be fed into an HFSS model of the waveguide, which can then return the expected electric field magnitude. The geometry of the waveguide is relatively simple, so modeling estimates are straightforward.  Crucially, this calculation represents the E-field strength outside of the vapor cell, but not necessarily inside of the cell. Vapor cells exhibit different levels of internal field attenuation based on cell material, vapor species, atom-surface interactions, and carrier frequency.  Frequency-dependent discrepancies between inner and outer field strengths tell us valuable information about the performance capabilities of different vapor cell materials. More information on these calibration methods and field discrepancies can be found in \citep{danielTEMpaper} and \citep{kayim2026}.

In the second phase, the coupler frequency lockpoint and LO field strength are simultaneously optimized. This is done by imposing heterodyne beat note conditions (strong LO field, weak signal (SIG) field, and a small frequency offset $\delta$ between them) while scanning the coupler laser frequency across the EIT resonance. For a fixed SIG field strength, the LO is varied in strength, and probe transmission is measured across the full coupler frequency scan. As the coupler scans, beat note signals appear in the probe transmission. The beat note frequency (50~kHz) is much faster than the coupler scan frequency (10~Hz), so these beat notes appear as fast oscillations in the probe transmission. The probe transmission signal is then demodulated by a lock-in amplifier referenced to the beat note frequency, with the resultant signal as the envelope magnitude of the beat note in optical frequency space. Because our coupler frequency scan is well-calibrated by an RF source, we can then reliably find the coupler frequency lockpoint that yields the strongest beat note amplitude for a given LO strength. We execute this routine over a range of LO strengths to find the optimal LO strength and coupler frequency combination (see Fig.~\ref{fig:optimize_singleShot}).

In the third and final phase, we calculate the sensitivity of our setup. In this paper, as well as many others, sensitivity is defined as the noise floor of the power spectral density~(PSD) of the photodiode output, scaled to electric field units (V/m / $\sqrt{\text{Hz}})$. To measure sensitivity, we lock laser frequencies, impose heterodyne conditions, and measure the resulting beat note in the probe transmission. To get the electric field magnitude at the atoms, we calculate the power spectrum of the beat note and determine the field strength using the measured rms voltage of the beat note and the known calibration. Once the field strength is determined, we convert the beat note into a field-scaled PSD and measure the noise floor around the beat note peak (see Fig.~\ref{fig:sensitivity_single}). As we will describe more in Sec.~\ref{sec:results}, we take sensitivity measurements with a wide range of SIG field strengths, which (1) gives us useful statistics on sensitivity uncertainty and (2) reveals field regimes at which heterodyne beat note detection is and is not valid (i.e.~determining the dynamic range of the receiver).

Photon shot noise (PSN) stands as a current fundamental limitation for Rydberg electric field sensors.  Photons striking a photodetector will generate electrons in a Poissonian distribution given by:

$$P(N,\Phi,T) = \frac{e^{-\Phi T}(\Phi T)^{N}}{N!}$$
where $N$ is the number of photons detected in time T and $\Phi$ is the mean photon flux.  It follows that in time T we expect a mean number of photons $\langle N \rangle = \Phi T$.  The arrival rate can then be calculated from the energy of the photons ($h \nu)$ and total optical power ($P_{opt}$) delivered to the photodiode: $\Phi = \frac{P_{opt}}{h \nu}$.  Notably, for a Poissonian distribution the variance is equal to the mean and the PSD ($S_{p}(f)$) is white (constant) in the frequency domain:

$$S_{p}(f) = (hf)^{2} \cdot 2\Phi$$
$$S_{p}(f) = 2 h f P_{opt} $$
\begin{equation}
    \Rightarrow PSN = \sqrt{2hfP_{opt}}
    \label{eq:PSN_eq}
\end{equation}
where the PSN is given in units of optical Watts in a bandwidth window ($W_{opt}/\sqrt{Hz}$) \cite{Saleh:1084451}.  To benchmark sensitivities to the PSN, it is convenient to convert from optical to electrical power using the responsivity and gain of the photodetector, and then scale to electric field units ($V/m/\sqrt{Hz}$).


The Python-based control suite Labscript \citep{starkey2013labscript} has greatly automated these three experimental control routines and analysis routines.  While these experimental routines are bespoke, tailored to communicate with our specific equipment generally through standard command for programmable interfaces, our analytical routines maintain a general applicability. Recognizing the broader potential for these analytical tools, these tools are available for public access on GitLab for the benefit of the scientific community \cite{github_needed}.

\section{Results}
\label{sec:results}

\subsection{Electric Field Calibration}
\label{subsec:calibration}
\begin{figure}[H]
    \centering
    \includegraphics[width=10 cm]{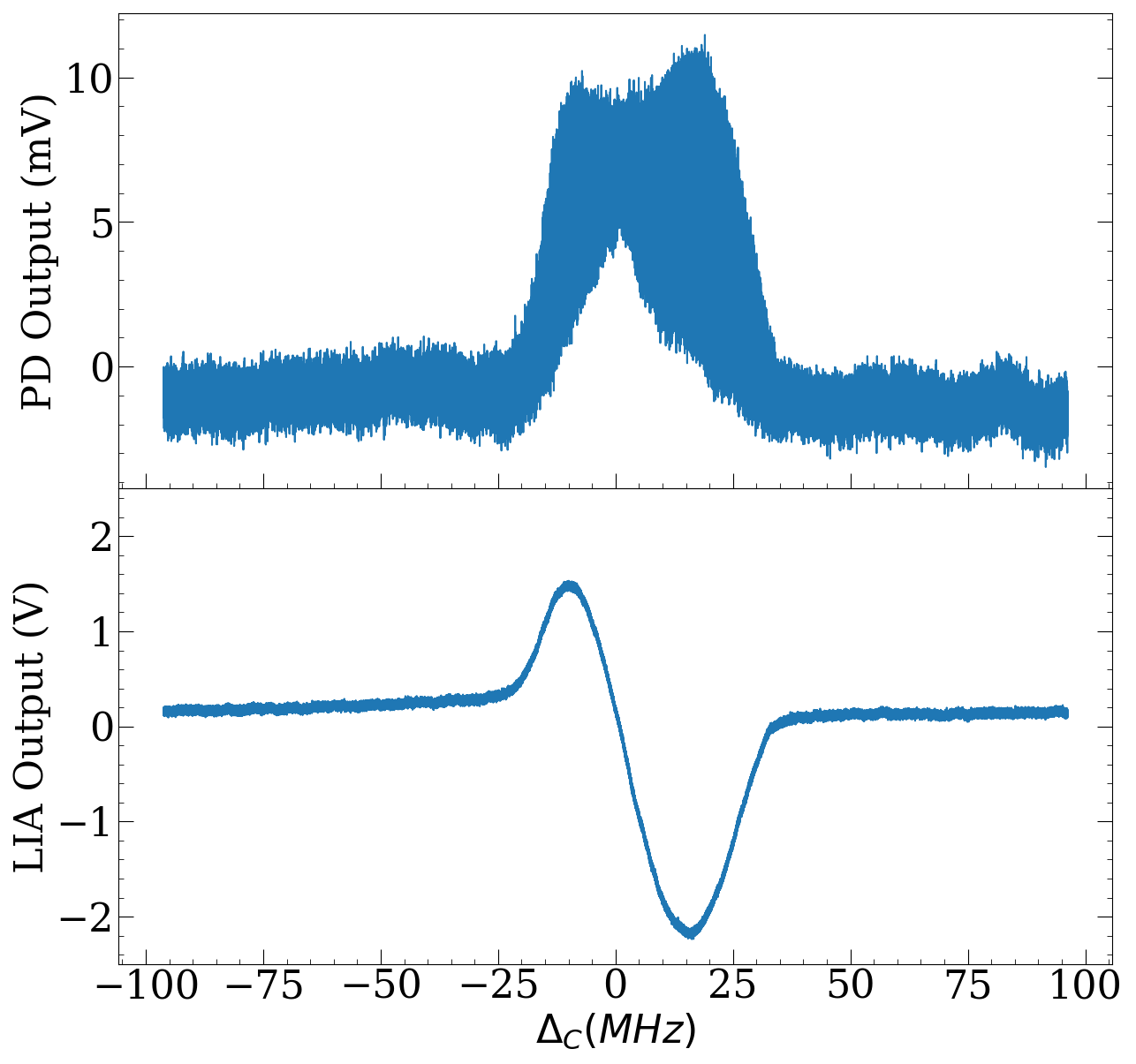}
    \caption{(Top)  EIT feature with the laser current modulated at 50 kHz.  (Bottom)  Lock-in amplifier (LIA) demodulation of the top figure.  Bottom axis is coupler frequency detuning from the center of the biased resonance.}
    \label{fig:dither}
\end{figure}

For each carrier frequency, we start with an atom-based calibration of the electric field. All of our chosen carrier frequencies are in the AC Stark regime (off-resonant detection), so calibration consists of measuring the AC Stark shift of the EIT line for a range of RF powers. As mentioned in Sec.~\ref{sec:materials}, we frequency-modulate the coupler laser as it scans across the EIT line. Demodulation of this dithered EIT line produces a dispersive signal with a zero crossing at the maximum of the EIT peak. We have found that zero crossings are easier to track than EIT maxima and yield cleaner calibrations as a result. Fig.~\ref{fig:dither} shows a modulated EIT signal (top plot), with dither frequency of 50~kHz and sweep frequency of 10~Hz. This trace is sent into a lock-in amplifier referenced to the dither frequency, and the in-phase (X) output (bottom plot) produces a zero crossing corresponding to the location of the EIT maximum.

\begin{figure}[H]
    \centering
    \includegraphics[width = 10 cm]{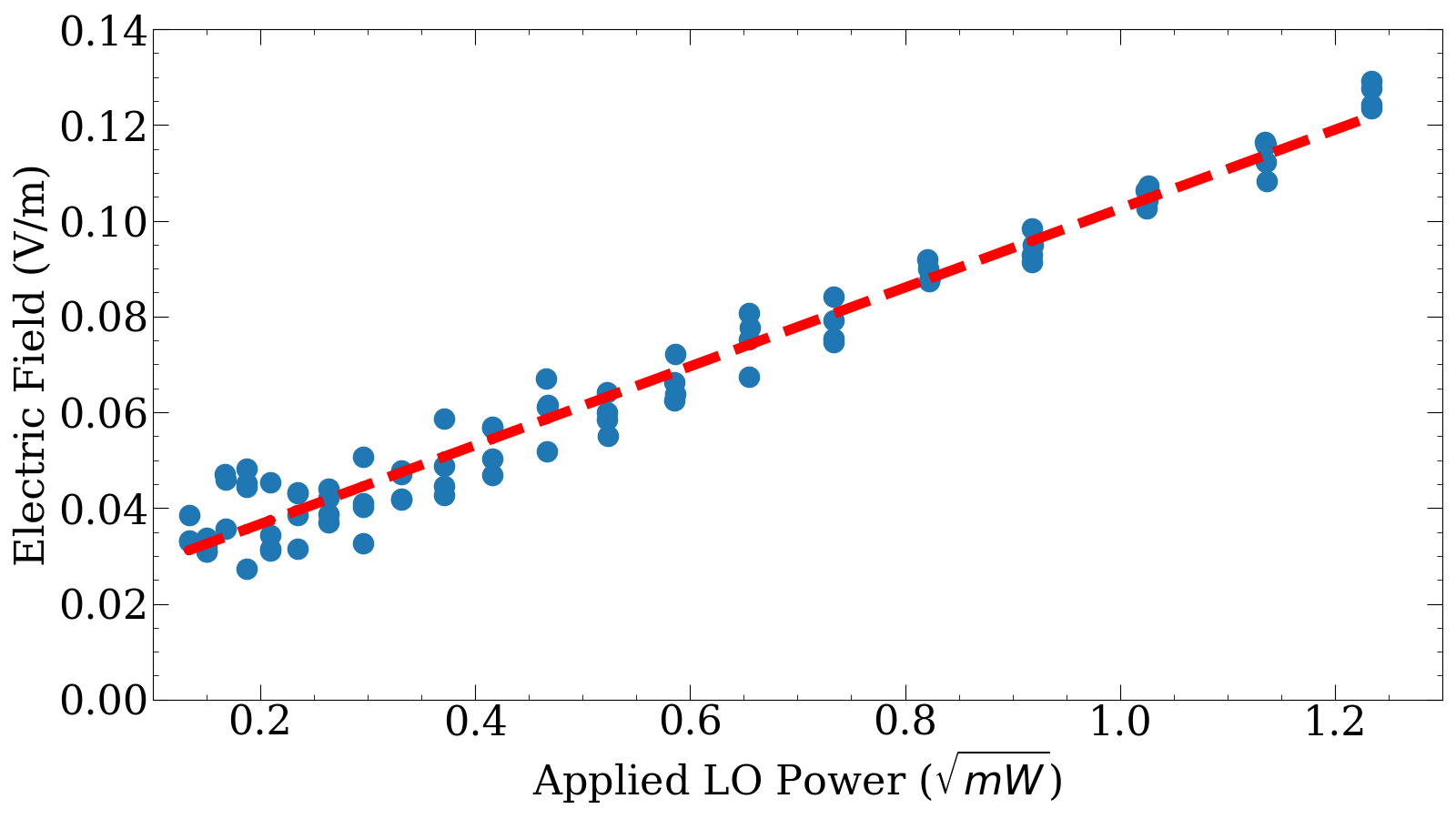}
    \caption{Electric field as calibrated by the atoms as a function of applied RF power to the TEM waveguide.  The dotted red line shows the linear relationship between the electric field as measured by the atoms and the square root of the power applied.}
    \label{fig:calibration}
\end{figure}

For each individual measurement, the location of the zero crossing is recorded with respect to the center of the frequency scan. This frequency offset is then converted to an electric field value using Eq.~\ref{eq:ACStarkShift}. Once electric fields have been calculated for every measurement, the field values are plotted against the square root of the applied RF power, which, after passing through the waveguide, is captured by a spectrum analyzer (Fig.~\ref{fig:calibration}). The field calibration is then obtained by applying a linear fit to the field measurements to determine the slope ($\frac{V/m}{\sqrt{P_{RF}}}$). The y-intercept of the fit is ignored, since the zero crossing measurement contains an arbitrary frequency offset. Only the relative shifts at changing RF powers matter for determining the calibration.

It should be noted that these calibrations represent the electric field inside of the vapor cell, as seen by the atoms. We have chosen the sapphire cell for these measurements because the sapphire vapor cell do not strongly shield HF/VHF signals from the atoms, unlike quartz or borosilicate glass. However, there is still a non-negligible amount of field screening, so the field measured inside of the cell is not necessarily indicative of the field outside of the cell. If Rydberg sensors are to be used in practical applications, this field screening must be accounted for, since the field properties outside of the cell are what end users ultimately care about. To that end, we have a second calibration method for the setup that is independent of atomic measurements. This calibration method, which is entirely classical, involves measuring S-matrix parameters for the vapor cell/waveguide assembly, building a corresponding 3D electromagnetic simulation, and then using this information to develop a model that can determine electric field strength for any RF signal of arbitrary frequency and throughput power. More information on this model can be found in \citep{danielTEMpaper} and \citep{kayim2026}. These electric field calibrations can then be compared with atom-based calibrations inside the vapor cell to determine frequency-dependent field screening.

\subsection{LO and Coupler Laser Optimization}
\label{subsec:optimization}
\begin{figure}[H]
    \centering
    \includegraphics[width=10 cm]{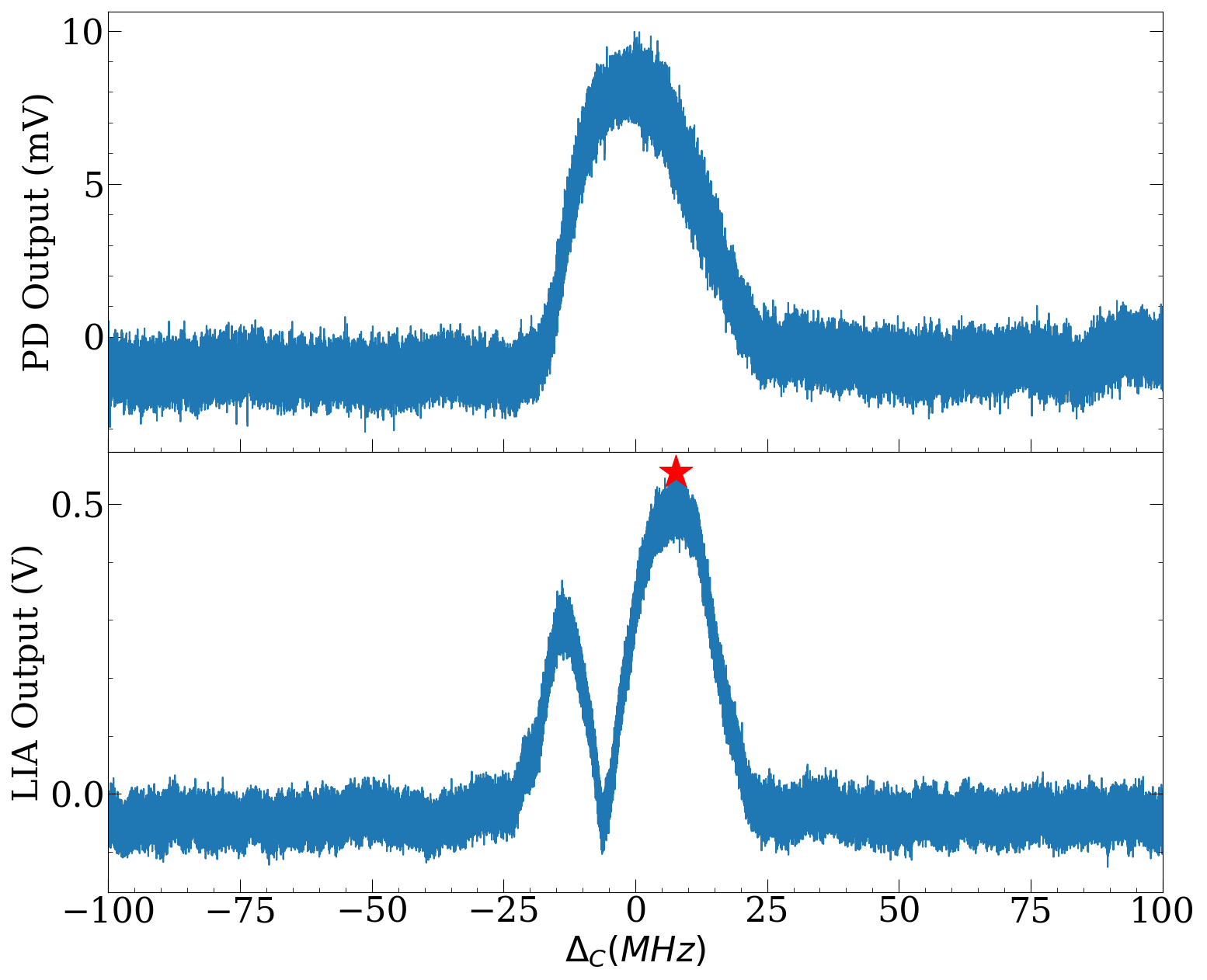}
    \caption{(Top) EIT biased by a strong local oscillator with a 50 kHz heterodyne beatnote imprinted on the feature. (Bottom)  Lock-in amplifier (LIA) demodulation of the top figure. Bottom axis is coupler frequency detuning from the center of the biased resonance.}
    \label{fig:optimize_singleShot}
\end{figure}

After calibrations are obtained, we optimize the heterodyne beat note signal on the LO power and 1260~nm laser frequency lockpoint for each carrier frequency. For these measurements, we disable the dither in the 1260~nm laser frequency, and enable heterodyne beat note detection with two RF fields: a stronger LO and the weaker, +50~kHz-detuned SIG. Using constant SIG power, we step the LO power and record the corresponding EIT trace. Fig.~\ref{fig:optimize_singleShot} shows a sample measurement: the top plot is an EIT peak with heterodyne conditions enabled. As the 1260~nm laser sweeps across the three-photon resonance, a beat note appears in the trace. The beat note varies in strength depending on the location of the sweep. This signal is sent into the lock-in amp, which is referenced to the beat note frequency of 50~kHz. The lock-in amp takes this signal and outputs a quasi-DC signal where the height corresponds to the beat note strength. This can be readily seen in the magnitude (R) output of the lock-in amp (bottom plot of Fig.~\ref{fig:optimize_singleShot}). For each experimental shot, we find the maximum of the lock-in amp output, along with the coupler frequency detuning where this maximum occurs.

\begin{figure}[H]
    \centering
    \includegraphics[width = 10 cm]{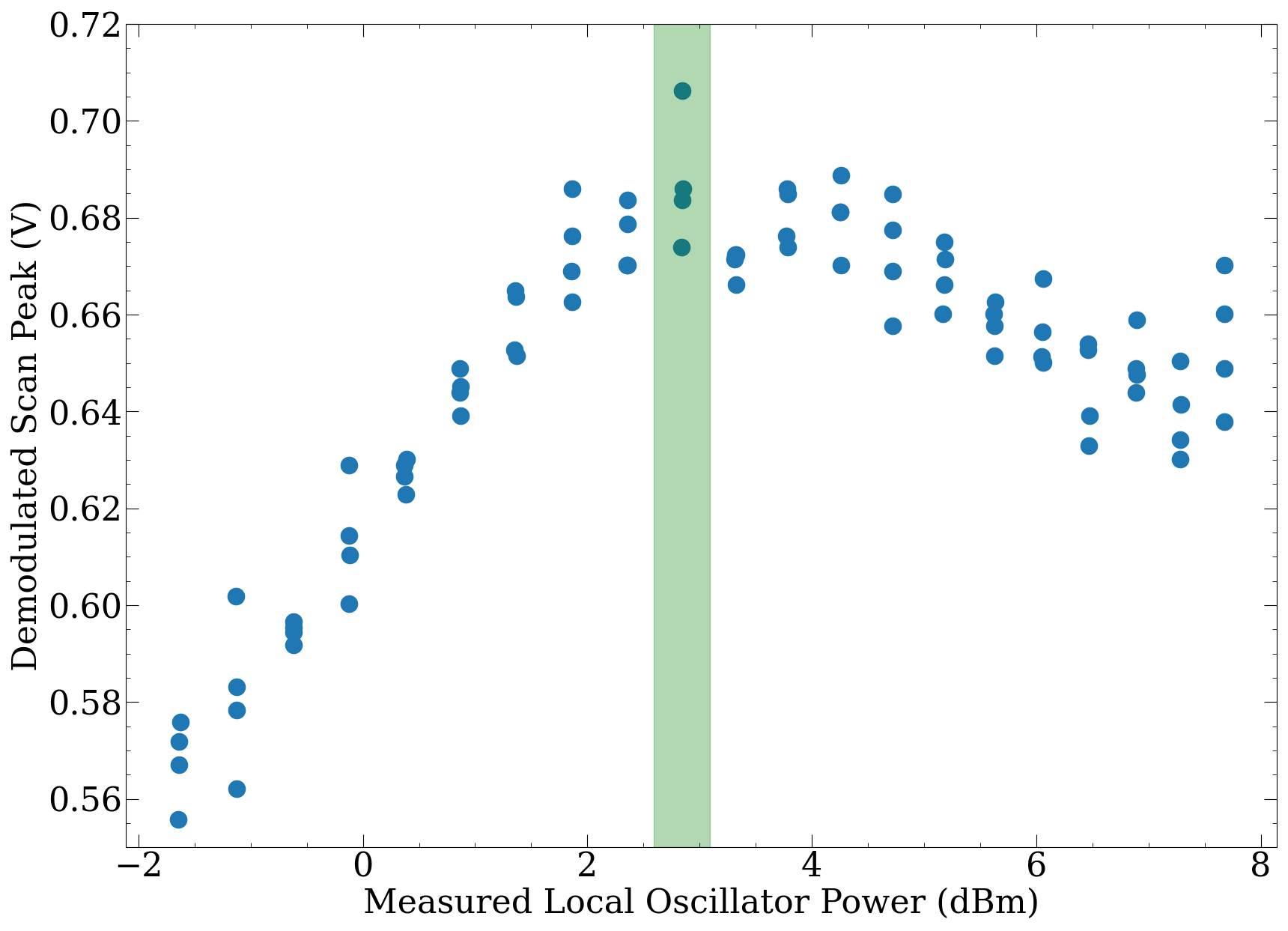}
    \caption{Maximum of demodulated EIT (Fig.~\ref{fig:optimize_singleShot} bottom) for a constant signal power as a function of applied local oscillator power.}
    \label{fig:optimization}
\end{figure}

Figure~\ref{fig:optimization} shows a plot of these measured maxima over a range of LO powers. At first the maximum beat note strength monotonically grows with LO power. After a certain point, the beat note strength rolls off and decreases with increasing LO power as the EIT lineshape broadens. For this particular set of measurements with carrier frequency of 40.68~MHz, the maximum beat note strength occurs at a measured LO power of $\sim$3~dBm (highlighted in green). Once we have determined the best LO operating point, we can go back to the single shot measurements and extract the value of the coupler detuning $\Delta_c$ where this maximum occurs. Thus, this method provides experimentally determined and optimized operating points for both LO field strength and 1260~nm laser frequency.

The ideal theoretical LO field strength for off-resonant RF detection is a question that is outside the scope of this paper, as it is a challenging problem compared to the ideal LO field strength for on-resonant detection. In the case of on-resonant detection with a Rydberg-Rydberg transition and assuming a perfectly homogenous LO, the optimal LO field strength is $E = \frac{2 \pi \hbar \Gamma_\text{EIT}}{d_\text{RR}} $\citep{jing2020heterodyne}. Here, $\hbar$ is the reduced Planck constant, $\Gamma_\text{EIT}$ is the EIT linewidth, and $d_\text{RR}$ is the transition dipole moment of the Rydberg-Rydberg transition. Thus, the optimal electric field strength for on-resonant detection is such that the transition Rabi rate, and therefore Autler-Townes splitting, is equivalent to the EIT linewidth. Futhermore, the optimal coupler laser frequency lockpoint is the saddle point between the two split peaks. In practice, the optimal LO used in off-resonant detection is dependent on many parameters, including principal quantum number, carrier frequency, and dynamic polarizabilities of the different Rydberg $J$-substates.

\subsection{Rydberg Sensor Sensitivity and Photon Shot-Noise Limit}
\label{subsec:sensitivity}
\begin{figure}[H]
    \centering
    \includegraphics[width=10 cm]{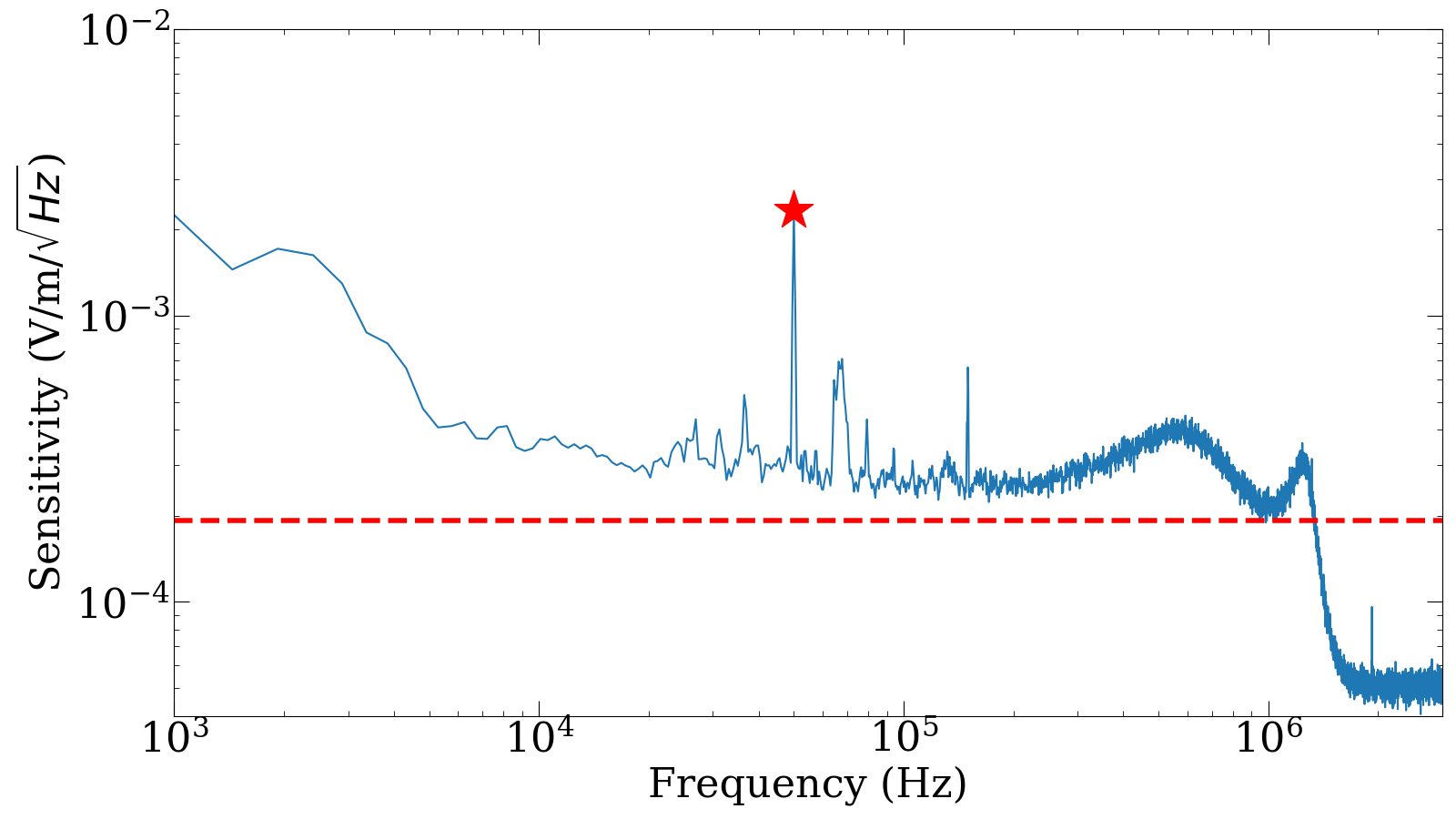}
    \caption{Calibrated electric field-scaled power spectral density.}
    \label{fig:sensitivity_single}
\end{figure}

After determining the optimal operating points for the LO power and coupler laser frequency by optimizing the heterodyne beatnote for a given signal field, we measured electric field sensitivity. For this measurement, the coupler laser was locked to the optimal frequency, and heterodyne detection was enabled with a beat note frequency of $\delta=50$~kHz. For a given SIG power, the beat note was measured, and then converted into a power spectrum and a PSD. Electric field strength was determined from the power spectrum and calibration from Sec.~\ref{subsec:calibration}. The PSD was then scaled to field units, and the noise floor around the beat note was measured and reported as the sensitivity, in units of ($\mu$V/m)/$\sqrt{\text{Hz}}$. Fig.~\ref{fig:sensitivity_single} shows an example of a field-scaled PSD with a beatnote at 50 kHz (red star), netting a sensitivity of roughly 300~($\mu$V/m)/$\sqrt{\text{Hz}}$. The strong roll-off in this trace is due to an analog 1~MHz low-pass filter to remove aliasing.

The PSN represents a fundamental sensitivity limit, providing a benchmark for system optimization. Consequently, PSN offers a rigorous standard for evaluating the performance of an EIT setup. Spectra were taken with the photodiode disconnected from the oscilloscope, as well as with no light striking the photodetector, to ensure electronic noise did not dominate the measurements.  Furthermore, off-resonance spectra were taken (not shown) with the probe laser far detuned (GHz) from the atomic transition (much greater than the $\sim$500 MHz Doppler-broadened linewidth).  During this measurement, it was observed that noise remained at the PSN level up to the analog filter's roll-off, as expected for simple light detection at the photodiode.  

To determine the total optical power at the photodetector, one can either measure it directly using an optical power meter or calculate it by scaling the photodiode's output voltage to the corresponding optical power. In this study, the first method was employed. The resulting PSN level (converted to field units) was calculated using Eq.~\ref{eq:PSN_eq} and is shown in Fig.~\ref{fig:sensitivity_single} as a dashed red line.  Below 10 kHz, significant noise was observed, primarily attributed to transit noise caused by atoms entering and exiting the laser beam. Beyond 10 kHz, the noise approached the PSN level but did not reach it until approximately 1 MHz.  The spectrum exceeding the PSN at all frequencies out to 1 MHz is clearly due to atom-light interactions \cite{Aoki}. As Rydberg-atom electric field sensing at the PSN is seemingly absent from the literature, achieving or surpassing this noise limit would set a new benchmark for Rydberg electrometers.  The addition of the dressing and coupling lasers was found to have a negligible impact on the noise spectrum.

\begin{figure}[H]
    \centering
    \includegraphics[width = 10 cm]{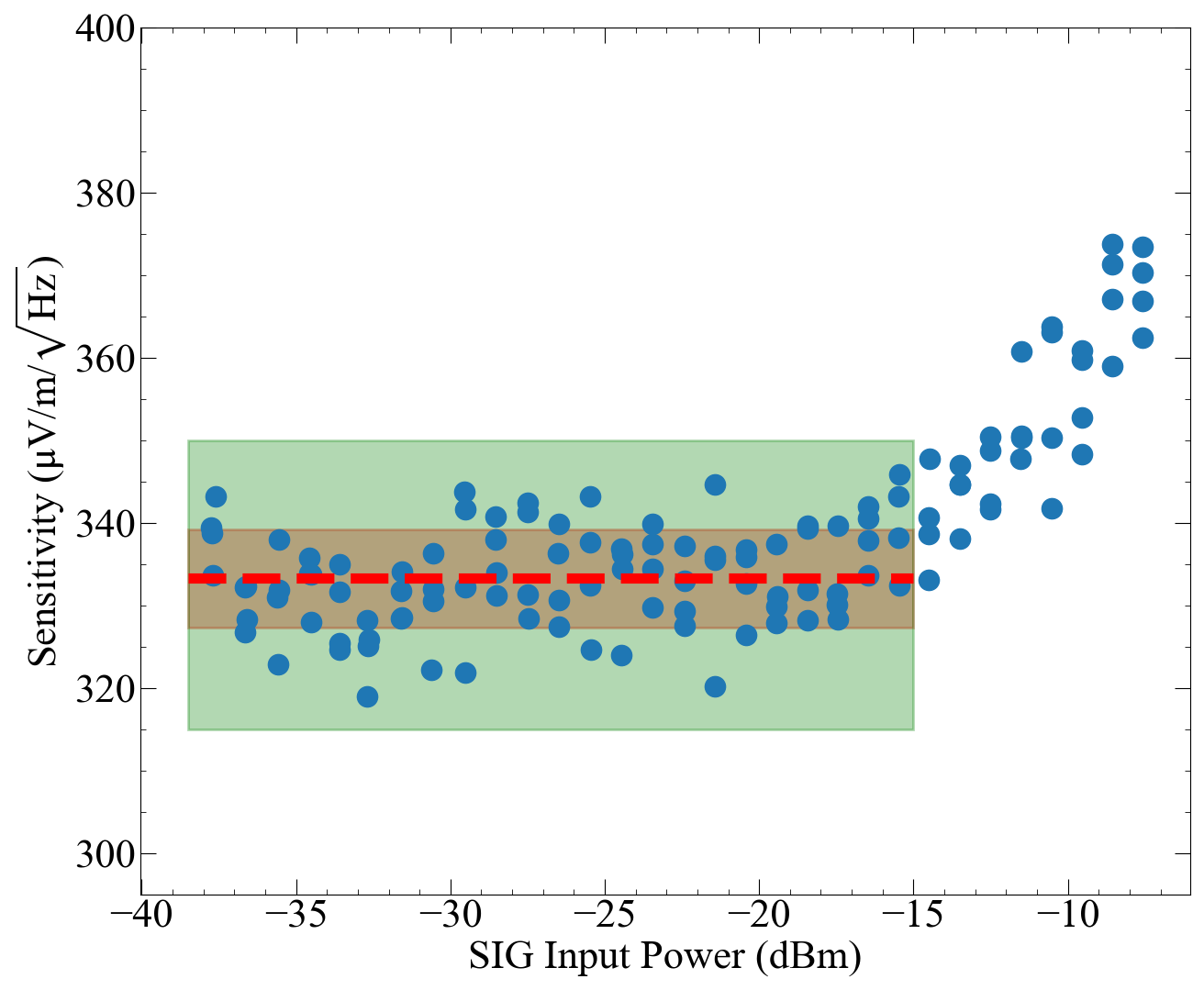}
    \caption{Measured sensitivity (as in Fig.~\ref{fig:sensitivity_single}) for a fixed, optimized coupler frequency and local oscillator power as a function of applied signal power.}
    \label{fig:sensitivity_multi}
\end{figure}

Heterodyne beatnotes were measured, and their PSDs were calculated across a range of SIG powers. This was done to gather useful statistics for sensitivity measurements and to investigate how the noise floor and sensitivity behave over a wide range of SIG powers. Fig.~\ref{fig:sensitivity_multi} presents sensitivity measurements for various measured SIG powers. While statistical variations exist between sensitivity measurements at equivalent SIG powers, a range is identified—highlighted in green—where sensitivity remains roughly constant despite variations in applied field strength. This stability suggests that the noise floor is relatively uniform within this regime. Outside this acceptable range, sensitivity degrades in magnitude as the SIG power deviates further from the stable regime. At very low SIG, the beatnote signal falls below the noise floor.  Conversely, at the high end, the SIG power becomes strong enough to violate heterodyne conditions ($E_\text{LO} >> E_\text{SIG}$), thereby compromising the measurement. Sensitivity measurements in both of these regimes are excluded when calculating the average sensitivity of the sensors (those points beyond the green box).  For carrier frequency of 40.68~MHz, we report a sensitivity of 352.2$\pm$13.7~($\mu$V/m)/$\sqrt{\text{Hz}}$ (average and standard deviation shown in dashed red line and highlighted in red, respectively). More sensitivities are listed in Tab.~\ref{tab:sensitivities}.

\begin{table}[H]
    \centering
    \begin{tabular}{c|c}
        Carrier Frequency [MHz] & Sensitivity [($\mu$V/m)/$\sqrt{\text{Hz}}$] \\
        \hline
        40.68 & 333 (6) \\
        27.12 & 449 (8) \\
        13.56 & 399 (10) \\
        6.78 & 125 (4)
    \end{tabular}
    \caption{Sensitivities for different ISM carrier frequencies.}
    \label{tab:sensitivities}
\end{table}

\section{Discussion}

We have demonstrated that Rydberg atom-based, off-resonant measurements of VHF-band signals can be calibrated and optimized using highly-automated routines. Our parameter optimization script effectively simplifies the otherwise time-intensive manual process of sweeping through LO bias field strengths and coupler laser optical detunings. This capability has high utility for potential Rydberg sensors in fielded applications, where rapid calibration and robust performance are crucial. Applying a powerful, low-noise, and spatially uniform local oscillator to the atoms along the beamline is critical in achieving peak performance in minimum detectable fields and sensitivities. Accordingly, the determination of optimal LO bias strength and the corresponding coupler frequency is an operational requirement.

While local oscillator field strength and coupler frequency are critical, they represent just a fraction of the broader parameter space within three-photon EIT heterodyne beatnote detection. Several other variables, including all laser parameters and environmental conditions, have substantial impacts on the performance of Rydberg-atom electric field sensors. For instance, laser intensities, linewidth, and waist sizes are consequential. Fine control of probe laser power is particularly vital for producing a high-contrast EIT signal, which enhances signal-to-noise ratio and enables better sensitivities. However, this improvement must be balanced against the trade-off introduced by photon shot noise, which increases as the square root of the probe power. The laser beam waist lends itself to the sensor's bandwidth through its effect on atom transit noise.  The operational temperature of the vapor cell represents another parameter that lacks a straightforward optimal setting. An elevated temperature increases atomic vapor density, resulting in stronger probe absorption and potentially enhancing signal strength. Low temperature leads to a lower atomic density and therefore fewer atoms detect the external field for a given laser waist, thereby scaling down the sensor's sensitivity.  These issues highlight ongoing challenges in Rydberg electrometer research.

In summary, while our work has advanced the parameter optimization of Rydberg sensors, it also underscores the complexity of achieving optimal sensor performance across a multidimensional parameter space. This complexity necessitates continuous development of optimization procedures not only for LO and coupler settings but also for fine-tuning other critical parameters such as laser power, beam waist, and cell temperature. Future iterations of Rydberg sensors will need these advancements to balance sensitivity, bandwidth, and practicality if they are to compete with conventional technologies in fielded applications. By addressing these challenges, Rydberg sensors hold the potential to become revolutionary tools in radio frequency metrology, sensing, and communications.





\vspace{6pt} 






\funding{This work was supported in part by Georgia Tech Research Institute Independent Research and Development (IRAD) funding.}

\dataavailability{Data available upon request.}

\conflictsofinterest{The authors declare no conflicts of interest.}



\abbreviations{Abbreviations}{
The following abbreviations are used in this manuscript:
\\

\noindent 
\begin{tabular}{@{}ll}
EIT & Electromagnetically induced transparency\\
RF & Radio Frequency\\
LO & Local oscillator\\
SWaP & Size, weight, and power\\
EOM & Electro-optic modulator\\
ISM & Industrial, scientific, and medical\\
PSD & Power spectral density\\
PSN & Photon shot noise
\end{tabular}
}




\isPreprints{}{
} 

\reftitle{References}


\bibliography{sensorsBib}

\PublishersNote{}
\end{document}